\documentclass[
reprint,
 amsmath,amssymb,
 aps,
]{revtex4-2}

\usepackage{hyperref}% add hypertext capabilities

\usepackage{graphicx}
\usepackage{siunitx}

\begin{document}

\title{Extremely weak electron-phonon coupling in Josephson junctions built on InAs on Insulator}
% Force line breaks with \\ 
\author{Giorgio De Simoni}
\email{giorgio.desimoni@nano.cnr.it}
\affiliation{NEST, Istituto Nanoscienze-CNR and Scuola Normale Superiore, Piazza S. Silvestro 12, I-56127 Pisa, Italy}

\author{Sebastiano Battisti}
\affiliation{NEST, Istituto Nanoscienze-CNR and Scuola Normale Superiore, Piazza S. Silvestro 12, I-56127 Pisa, Italy}

\author{Alessandro Paghi}
\affiliation{NEST, Istituto Nanoscienze-CNR and Scuola Normale Superiore, Piazza S. Silvestro 12, I-56127 Pisa, Italy}

\author{Lucia Sorba}
\affiliation{NEST, Istituto Nanoscienze-CNR and Scuola Normale Superiore, Piazza S. Silvestro 12, I-56127 Pisa, Italy}

\author{Francesco Giazotto}
\affiliation{NEST, Istituto Nanoscienze-CNR and Scuola Normale Superiore, Piazza S. Silvestro 12, I-56127 Pisa, Italy}

%\date{\today}

\begin{abstract}
InAs‐ on‑Insulator (InAsOI) enables an effective superconducting proximity effect and extremely weak electron–phonon (e-ph) coupling, allowing precise electronic‑temperature control with minimal power. Using Josephson junction thermometry, we extract sub‑Kelvin e–ph coupling parameters, confirming strong thermal decoupling and robust superconducting performance. The combination of weak e–ph interaction and full electrostatic tunability makes InAsOI a powerful platform for coherent caloritronics, ultrasensitive bolometry, single‑photon detection, and gate‑controlled superconducting thermal circuits. 
\end{abstract}

\maketitle

\section{Introduction}
\begin{figure*}[t]
  %\centering
  
  \includegraphics[width=\linewidth]{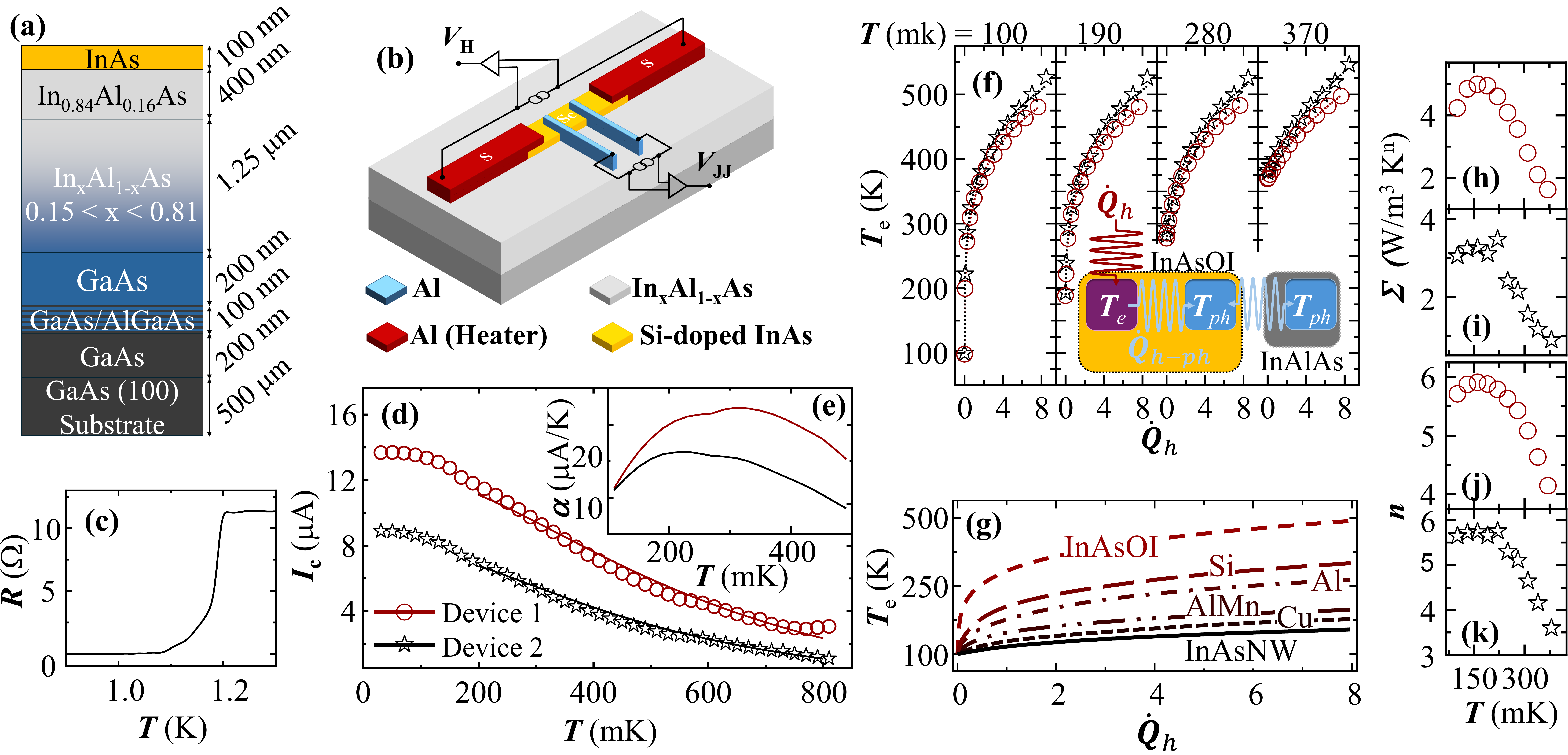}
  \caption{\textbf{InAsOI as a caloritronic material.}
  (a) Layer structure of InAsOI
  (b) Scheme of the InAsOI mesa (yellow) with Al Joule heaters (red) and Al JJ leads (blue). The device wiring scheme is also represented.
  (c) JJ resistance \emph{vs} bath temperature $T$ showing a transition near the Al critical temperature $T_{Al}\simeq$\SI{1.2}{\kelvin}.
  (d) Critical current ($I_c$) of the Josephson junction as a function of bath temperature $T$ for both devices. The solid lines are the best fit of $I_c(T)$ against the model describing long diffusive Josephson junctions in the high‑temperature limit.
  (e) Temperature responsivity $\alpha$ of the Josephson thermometers, as a function of $T$ for both the devices.
  (f) Experimental data (scatter plot) and the fit (dashed lines) using \autoref{eq1} at selected bath temperatures for $T_e$ as a function of the injected power $\dot Q_{h}$. \textit{Inset}: A scheme of the predominant heat currents flowing through the InAsOI system: $\dot Q_{h}$ is the injected power from the Joule heaters, whereas $\dot Q_{e_ph}$ represents the power dissipated to the InAsOI lattice phonons which we assume to be perfectly coupled to that of the substrate and residing at the temperature of the cryostat.
  (g)-(h): The value of $\Sigma$ extracted from the fitting procedure as a function of $T$ for device 1 (g) and 2 (h).
  (i)-(k): The value of $n$ extracted from the fitting procedure as a function of $T$ for device 1 (i) and 2 (j).
  }
  \label{fig:device}
\end{figure*}

Hybrid superconductor–semiconductor (S–Sc) platforms enable fast, low‑dissipation cryogenic electronics. Among III–V materials, InAs is particularly advantageous because its surface electron accumulation suppresses Schottky barriers, yielding highly transparent S–Sc interfaces and strong proximity effects. InAs‑based Josephson junctions (JJs) and Josephson field‑effect transistors (JoFETs) have been implemented in 2DEGs, quantum wells, and nanowires, but each platform presents intrinsic limitations such as parasitic conduction, reduced induced‑gap, or limited scalability (see \cite{Paghi2025_AFM} and the references therein).
InAs-on-Insulator (InAsOI) \cite{Paghi2025_AFM, Senesi2025} is a recently introduced heterostructure [see \ref{fig:device}(a)] that overcomes these constraints by combining an exposed InAs surface with a step‑graded In$_x$Al$_{1-x}$As buffer grown on semi‑insulating GaAs, which becomes highly resistive at low temperature. Analogously to SOI in CMOS technology, InAsOI electrically isolates neighboring devices while preserving transparent contacts and enabling direct electrostatic gating of the InAs channel. Its carrier density is widely tunable, supporting precise control of mobility and mean free path. On this platform, planar Al/InAs/Al junctions achieve switching current densities comparable to the best Nb-based bulk-InAs devices and superior to those of quantum well- and nanowire-based implementations. InAsOI also supports efficient JoFET operation, including full suppression of the switching current and strong modulation of the normal‑state resistance\cite{Paghi2025_ACSAELM, Paghi2024_FerroelectricJoFET}, with low leakage—features relevant for gatemon qubits and superconducting multiplexers\cite{Paghi2024_SupercurrentMultiplexing}.
Here, we investigate the thermal properties of InAsOI, focusing on its ability to suppress electron–phonon (e–ph) coupling and decouple the electronic temperature from the phonon bath. Indeed, coherent caloritronics relies on controlling heat currents in regimes where e–ph coupling is weak, a condition that is naturally met in semiconductors but difficult to combine with the high‑quality superconducting proximity effect. InAsOI provides both: highly transparent S–Sc interfaces and intrinsically weak e–ph relaxation. As we show, very small heating powers are sufficient to drive the electron system out of equilibrium, demonstrating strong thermal decoupling. These characteristics make InAsOI an ideal platform for integrating Josephson devices into coherent caloritronic architectures. In the following, we quantify the e–ph coupling in this system and compare it with other hybrid S–Sc platforms.

\section{Methods}
\label{sec:methods}

\subsection{Heterostructure and device layout}
\label{subsec:stack}
Electron–phonon coupling is extracted from an S–Sc–S Josephson junction used as an electronic thermometer to obtain the electronic temperature $T_e$ under fixed Joule heating. Measuring $T_e$ at different bath temperatures $T$ then allows determining the e–ph relaxation via a power‑law fit following Wellstood’s method (see \cite{Battisti2024_APL} and references therein).
The devices are fabricated from an InAs‑on‑Insulator heterostructure containing a 100 nm-thick InAs epilayer \cite{Senesi2025} with a room temperature carrier density of $n_{3D} \sim 2.5 \times 10^{18}$ cm$^{-3}$, by the fabrication procedure described in \cite{Battisti2024_APL}.
The complete structure [Fig.~\ref{fig:device}(b)] consists of a $\sim$ 20 $\mu$m by 6 $\mu$m InAs mesa, two Joule heaters separated by $\sim17$ $\mu$, and Al JJ leads separated by $\sim 550$ nm.

\subsection{Cryogenic setup and thermometer calibration}
\label{subsec:cal}
The device wiring follows the layout in Fig.~\ref{fig:device}(b), and the measurements were carried out in a low-pass filtered dilution refrigerator with a base temperature of 20 mK. The junction resistance $R(T)$ for a representative sample [Fig.~\ref{fig:device}(c)] shows a superconducting transition at $T_c\simeq 1.2$ K, matching the critical temperature of the Al film $T_{Al}=1.196$ K, thus confirming an efficient proximitization. The critical current $I_c(T)$ for two nominally identical devices (labeled 1 and 2) is reported in Fig.~\ref{fig:device}(d). Since the heaters were off, $T=T_e$. Fitting these data to the long diffusive‑junction model yields effective junction lengths $L_1\simeq 1.43$ $\mu$m and $L_2 \simeq  1.71$ $\mu$m, both larger than the lithographic $\sim$ 550 nm interelectrode spacing, and consistent with diffusive transport and Cooper‑pair penetration beneath the Al leads.
The $I_c (T)$ curves serve as calibration functions for the JJ thermometers used to extract the electronic temperature of the InAs mesa. The thermometer operates optimally between $\sim 100$ and $\sim 200$ mK, where $I_c(T)$ is nearly linear, allowing a reliable inversion from $I_C$ to $ T_e$. Outside this range, the reduced slope limits sensitivity. The responsivity $\alpha=\partial I_c/\partial T$ [Fig.~\ref{fig:device}(e)] quantifies this, and together with the switching‑current noise $\delta I_C \simeq 30$ nA and our setup bandwidth $Bw\simeq300$ Hz, yields a noise‑equivalent temperature $NET \simeq 100 - 200$ $\mu\mathrm{K}/\sqrt{\mathrm{Hz}}$. This performance is comparable to that of state‑of‑the‑art superconducting thermometers and underscores the suitability of this platform for calorimetric and single‑photon‑detection applications.

\section{Assessment of the e-ph coupling constant}
\label{sec:ephcoupl}

The inset of Fig.~\ref{fig:device}(f) shows the main heat currents in the InAs mesa of volume $V$. The injected power $\dot Q_{h}=I_H V_H$ is provided by the current‑biased Joule heaters. Energy relaxation occurs through $e-ph$ coupling, described by the material‑dependent parameters $\Sigma$ and $n$, such that $\dot Q_{e-ph}=\Sigma V (T_e^n - T_{ph}^n)$. Since the Kapitza resistance between the InAs phonons and the InAlAs substrate is negligible, we assume $ T _ {ph} = T$. Heat leakage through heaters and JJ leads is ignored because, throughout the experiment, $k_B T_e \ll \Delta_{Al} \simeq180,\mu$eV, so all superconducting contacts act as ideal Andreev mirrors. Under steady‑state conditions, the heat balance yields
\begin{equation}
    T_e=\sqrt[n]{\frac{\dot Q_{inj}}{\Sigma V}+T^n},
    \label{eq1}
\end{equation}
$\Sigma$ and $n$ are obtained by fitting $T_e$ vs. $\dot Q_h$ at fixed bath temperatures [see Fig.~\ref{fig:device}(f) and Ref.~\cite{Battisti2024_APL}]. The fits reproduce the data extremely well, with the worst reduced chi‑square of $\chi_{\mathrm{red}}^2 \simeq 1.1\times10^{-4}$.
Figures~\ref{fig:device}(h,i) show that $\Sigma$ is essentially constant between 100 and 220 mK, with  $\Sigma_1=(4.2\pm0.4)\times10^7$,W/(m$^3$K$^n$) and $\Sigma_2=(3.0\pm0.3)\times10^7$,W/(m$^3$K$^n$) for devices 1 and 2, respectively. Over the same interval, the exponent also remains stable, yielding $n_1=5.84\pm0.14$ and $n_2=5.72\pm0.12$ [Figs.~\ref{fig:device}(j,k)]. Because the units of $\Sigma$ depend on $n$, a direct comparison between materials is made by plotting $T_e$ vs. $\dot Q_h$ at $T=100$ mK for identical mesa volumes [Fig.\ref{fig:device}(g)]. The InAsOI curve (device 1) is compared with the theoretical curves for other materials using Eq.~(\ref{eq1}).
This normalized comparison shows that InAsOI exhibits the highest increase in electronic temperature for the same injected power. 
Although the absolute increase may appear modest, the power‑law dependence implies differences in $\Sigma$ of several orders of magnitude. These results confirm that InAsOI possesses an exceptionally weak electron–phonon coupling while still enabling high‑performance Josephson junctions. The observation that $n\approx 6$ is consistent with dirty‑limit transport (see \cite{Battisti2024_APL} and the references therein), where the electron mean free path ($\approx 178$ nm) is much shorter than the phonon wavelength ($\approx 1$ $\mu$m), a regime known to yield a $T^6$ dependence. The decrease in $\Sigma$ and $n$  above 220 mK can be attributed to the temperature-driven reduction of the $e-ph$ scattering length, which becomes comparable to the mesa length around $T_e\approx 380$ mK, invalidating the assumption of uniform temperature across the mesa, thus limiting the applicability of the power‑law model\cite{Battisti2024_APL}.

\section{Conclusions and perspectives}
\label{sec:conclusions}
In summary, we demonstrated that Al/InAsOI hybrid structures combine exceptionally weak $e–ph$ coupling with robust superconducting proximitization, outperforming metals and matching or surpassing other semiconductor platforms. This makes InAsOI an ideal candidate for caloritronic devices, for ultrasensitive bolometry and photon detection. Crucially, the transport properties of InAsOI can be tuned electrostatically, thus introducing a powerful alternative to traditional magnetic‑flux control, enabling gate‑driven caloritronic elements where thermal currents can be modulated via a gate voltage. This added degree of freedom broadens the scope of caloritronics and opens the path toward gate‑controlled thermal circuits\cite{Battisti2025_PhotonicHeatTransistor}.

\end{document}